\documentclass[11pt,english,onecolumn]{IEEEtran}
\usepackage[T1]{fontenc}
\usepackage[latin9]{inputenc}
\usepackage{amsthm}
\usepackage{amsmath}
\usepackage{amssymb}
\usepackage{mathdots}
\usepackage{graphicx}
\usepackage{setspace}
\PassOptionsToPackage{version=3}{mhchem}
\usepackage{mhchem}
\usepackage{esint}
\makeatletter
\theoremstyle{plain}
\newtheorem{thm}{\protect\theoremname}
\theoremstyle{plain}
\newtheorem{prop}[thm]{\protect\propositionname}
\theoremstyle{plain}
\newtheorem{cor}[thm]{\protect\corollaryname}
\theoremstyle{remark}
\newtheorem{rem}[thm]{\protect\remarkname}
\theoremstyle{definition}
\newtheorem{defn}[thm]{\protect\definitionname}
\theoremstyle{plain}
\newtheorem{lem}[thm]{\protect\lemmaname}

\usepackage{hyperref}
\usepackage{breqn}

\allowdisplaybreaks

\newcommand{\dfn}{\triangleq}

\newcommand{\reals}{{\rm I\!R}}

\global\long\def\P{\mathbb{P}}
\global\long\def\E{\mathbb{E}}
\global\long\def\I{\mathbb{I}}

\global\long\def\EE{E}

\global\long\def\s[#1]{\textnormal{\scriptsize #1}}

\global\long\def\trre[#1,#2]{\overset{{\scriptstyle (#2)}}{#1}} % transition explained

\author{
\authorblockN{Nir Weinberger and Neri Merhav}

\authorblockA{Dept. of Electrical Engineering\\
     	    Technion - Israel Institute of Technology\\
Technion City, Haifa 3200004, Israel
} \\
\authorblockA{\{nirwein@tx, merhav@ee\}.technion.ac.il}
}
\makeatother

\usepackage{babel}
\providecommand{\corollaryname}{Corollary}
\providecommand{\definitionname}{Definition}
\providecommand{\lemmaname}{Lemma}
\providecommand{\propositionname}{Proposition}
\providecommand{\remarkname}{Remark}
\providecommand{\theoremname}{Theorem}

\begin{document}

\title{Simplified Erasure/List Decoding}
\maketitle
\begin{abstract}
We consider the problem of erasure/list decoding using certain classes
of simplified decoders. Specifically, we assume a class of erasure/list
decoders, such that a codeword is in the list if its likelihood is
larger than a threshold. This class of decoders both approximates
the optimal decoder of Forney, and also includes the following simplified
subclasses of decoding rules: The first is a function of the output
vector only, but not the codebook (which is most suitable for high
rates), and the second is a scaled version of the maximum likelihood
decoder (which is most suitable for low rates). We provide single-letter
expressions for the exact random coding exponents of any decoder in
these classes, operating over a discrete memoryless channel. For each
class of decoders, we find the optimal decoder within the class, in
the sense that it maximizes the erasure/list exponent, under a given
constraint on the error exponent. We establish the optimality of the
simplified decoders of the first and second kind for low and high
rates, respectively. \end{abstract}
\renewcommand\[{\begin{equation}}
\renewcommand\]{\end{equation}}
\renewenvironment{align*}{\align}{\endalign}

\section{Introduction}

An ordinary decoder must always decide on a single codeword, and consequently,
its decision regions form a partition of the space of output vectors.
In his seminal paper \cite{Forney68}, Forney defined a family of
generalized decoders with decision regions which do not necessarily
satisfy the above property. An \emph{erasure} decoder has the freedom
not to decode, and so, its decision regions are disjoint but not necessarily
cover of the space of channel output vectors. A \emph{list} decoder
may decide on more than one codeword, thus its decision regions may
overlap. An erasure/list decoder can practically be used in cases
where an additional mechanism is used to resolve the ambiguity left
after a non-decisive decoding instance. For example, if the transmitted
codewords at different times are somewhat dependent, then an outer
decoder may recover from an erasure by ``interpolating'' from neighboring
codewords, or eliminate all codewords in the list, but one. Such a
case is present in concatenated codes\emph{ }where the inner decoder
may be an erasure/list decoder. As another example, consider the case
where a feedback link may signal the transmitter on an erasure event.
In this case, the transmitter may send repeatedly the codeword until
there is no erasure. Indeed, this is the case in rateless codes \cite{Burnashev_feeddback,shulman2003communication},
where at each time instant, the decoder needs to decide whether to
continue acquiring additional output symbols, or decide on the transmitted
codeword. 

When considering list decoders, a distinction should be made between
list decoders with a fixed list size (see for example \cite{Neri_List},
and references therein), and decoders for which the list size is a
function of the output vector and thus a random variable. In the latter
case, a natural trade-off in designing the decoder is the error probability,
i.e., the probability that the correct codeword is not on the list,
and the average list size. In \cite{Forney68}, Forney has generalized
the Neyman-Pearson Lemma and derived the decoder which optimally trades
off between error probability and average list-size. This optimal
list decoder and the optimal erasure decoder (which optimally trades
off between error probability and erasure probability), were found
to have the same structure. 

In addition, Forney has extended the analysis of Gallager for ordinary
decoding \cite[Chapter 5]{gallager1968information}, and derived lower
bounds on the random coding error exponent and the list size exponent
(the normalized logarithm of the expected list size). More recently,
Somekh-Baruch and Merhav \cite{exact_erasure} have found exact expressions
for the random coding exponents, inspired by the statistical-mechanics
perspective on ensembles of random codes \cite{Neri_statistical_physics}.
Over the years, research was aimed towards various extensions. For
example, in \cite[Chapter 10]{Csiszar2011} \cite{Moulin_universal_erasure,universal_minimax_erasure,universal_erasure_exact}
universal versions of erasure/list decoders, i.e., decoders that perform
well even under channel uncertainty. In \cite{telatar1998exponential},
the size of the list was gauged by its $\rho$-th moment, where $\rho=1$
corresponds to the average list size, as in \cite{Forney68}, and
an achievable pair of error exponent and list size exponents was found. 

In essence, as we shall see, Forney's optimal erasure/list decoder
is conceptually simple - a codeword belongs to the output list if
its a posteriori probability, conditioned on the output vector, is
larger than a\emph{ fixed} threshold. However, as is well known, the
prospective implementation of such a decoder is notoriously difficult,
since it requires the calculation of the sum of likelihoods (c.f.
Subsection \ref{sub:System-Model-and}). These likelihoods typically
decrease exponentially as the block length increases, and thus have
a large dynamic range. Therefore, an exact computation of this sum
is usually not feasible. Moreover, the number of summands is equal
to the number of codewords and thus increases exponentially with the
block length. In fact, it was proposed already by Forney himself to
approximate this sum by its maximal element \cite[eq. (11a)]{Forney68}. 

It is the purpose of this paper to study classes of simplified decoders
which avoid the need to confront such numerical difficulties. To this
end, we first formulate the optimal erasure/list decoder as a threshold
decoder, and propose to replace the optimal threshold with a threshold
from a fairly general class. Any threshold from this class may represent
some complexity constraint of the decoder. We then consider two subclasses
of this class of decoders. For the first subclass, which is well suited
for high rates, the threshold function is extremely simple - it depends
on the output vector only. For the second subclass, which is well
suited for low rates, the threshold is the log-likelihood (plus an
additive fixed term), which is the approximation proposed by Forney.
Then, for a given discrete memoryless channel (DMC), we derive the
exact random coding exponents for the ensemble of fixed composition
codebooks, associated with decoders from the above general class and
its two subclasses. Next, for the general class of decoders, and for
its two subclasses, we find the optimum decoder, within the given
class, which achieves the largest list size exponent for a prescribed
error exponent. This enables the establishment of the optimality of
the simplified decoders of the first and second kind, for low and
high rates, respectively.

The outline of the rest of the paper is as follows. In Section \ref{sec:Problem-Formulation},
we establish notation conventions, provide some background of known
results, and define the relevant classes of decoders. In Section \ref{sec:Exponents-of-Threshold},
we derive the exact random coding exponents for all defined decoders,
and in Section \ref{sec:Optimal-Choice-of}, we find the optimal decoder
within each class. In Section \ref{sec:Optimality-of-Simplified},
we discuss the optimality of the various simplified decoders defined,
for low and high rates. Finally, in Section \ref{sec:Numerical-Example},
we compare the random coding exponents of the optimal decoder, as
obtained in \cite{exact_erasure}, with the exponents of the suboptimal
decoders, for simple numerical examples. All the proofs are deferred
to Appendix \ref{sec:Proofs}.

\section{Problem Formulation\label{sec:Problem-Formulation}}

\subsection{Notation Conventions\label{sub:Notation-Conventions}}

Throughout the paper, random variables will be denoted by capital
letters, specific values they may take will be denoted by the corresponding
lower case letters, and their alphabets, similarly as other sets,
will be denoted by calligraphic letters. Random vectors and their
realizations will be denoted, respectively, by capital letters and
the corresponding lower case letters, both in the bold face font.
Their alphabets will be superscripted by their dimensions. For example,
the random vector $\mathbf{X}=(X_{1},\ldots,X_{n})$, ($n$ - positive
integer) may take a specific vector value $\mathbf{x}=(x_{1},\ldots,x_{n})$
in ${\cal X}^{n}$, the $n$-th order Cartesian power of ${\cal X}$,
which is the alphabet of each component of this vector.

A joint distribution of a pair of random variables $(X,Y)$ on ${\cal X}\times{\cal Y}$,
the Cartesian product alphabet of ${\cal X}$ and ${\cal Y}$, will
be denoted by $Q_{XY}$ and similar forms, e.g. $\tilde{Q}_{XY}$.
Usually, we will abbreviate this notation by omitting the subscript
$XY$, and denote, e.g, $Q_{XY}$ by $Q$. The $X$-marginal (\textbf{$Y$}-marginal),
induced by $Q$ will be denoted by $Q_{X}$ (respectively, $Q_{Y}$),
and the conditional distributions will be denoted by $Q_{Y|X}$ and
$Q_{X|Y}$. In accordance with this notation, the joint distribution
induced by $Q_{X}$ and $Q_{Y|X}$ will be denoted by $Q_{X}\times Q_{Y|X}$,
i.e. $Q=Q_{X}\times Q_{Y|X}$. 

For a given vector $\mathbf{x}$, let $\hat{Q}_{\mathbf{x}}$ denote
the empirical distribution, that is, the vector $\{\hat{Q}_{\mathbf{x}}(x),~x\in{\cal X}\}$,
where $\hat{Q}_{\mathbf{x}}(x)$ is the relative frequency of the
letter $x$ in the vector $\mathbf{x}$. Let ${\cal T}_{P}$ denote
the type class associated with $P$, that is, the set of all sequences
$\{\mathbf{x}\}$ for which $\hat{Q}_{\mathbf{x}}=P$. Similarly,
for a pair of vectors $(\mathbf{x},\mathbf{y})$, the empirical joint
distribution will be denoted by $\hat{Q}_{\mathbf{xy}}$. 

The mutual information of a joint distribution $Q$ will be denoted
by $I(Q)$, where $Q$ may also be an empirical joint distribution.
The information divergence between $Q$ and $P$ will be denoted by
$D(Q\|P)$, and the conditional information divergence between the
empirical conditional distribution $Q_{Y|X}$ and $P_{Y|X}$, averaged
over $Q_{X}$, will be denoted by $D(Q_{Y|X}\|P_{Y|X}|Q_{X})$. Here
too, the distributions may be empirical. 

The probability of an event ${\cal A}$ will be denoted by $\P\{{\cal A}\}$,
and the expectation operator will be denoted by $\E\{\cdot\}$. Whenever
there is room for ambiguity, the underlying probability distribution
$Q$ will appear as a subscript, e.g., $\E_{Q}\{\cdot\}$. The indicator
function will be denoted by $\I\{\cdot\}$. Sets will normally be
denoted by calligraphic letters. The probability simplex over an alphabet
${\cal X}$ will be denoted by ${\cal S}({\cal X})$. The complement
of a set ${\cal A}$ will be denoted by $\overline{{\cal A}}$. Logarithms
and exponents will be understood to be taken to the natural base.
The notation $[t]_{+}$ will stand for $\max\{t,0\}$. For two positive
sequences, $\{a_{n}\}$ and $\{b_{n}\}$, the notation $a_{n}\doteq b_{n}$
will mean asymptotic equivalence in the exponential scale, that is,
$\lim_{n\to\infty}\frac{1}{n}\log(\frac{a_{n}}{b_{n}})=0$.

\subsection{System Model and Background\label{sub:System-Model-and}}

Consider a DMC, characterized by a finite input alphabet ${\cal X}$,
a finite output alphabet ${\cal Y}$, and a given matrix of single-letter
transition probabilities $\{W(y|x),~x\in{\cal X},~y\in{\cal Y}\}$.
Conditioning on the channel input $\mathbf{X}=\mathbf{x}$, the distribution
of the output $\mathbf{Y}$ is given by 
\begin{equation}
\P(\mathbf{Y=}\mathbf{y}|\mathbf{X}=\mathbf{x})=\prod_{i=1}^{n}W(y_{i}|x_{i}).\label{eq: memoryless conditional}
\end{equation}
We will use the shorthand notations $P(\mathbf{x},\mathbf{y})\dfn\P(\mathbf{X}=\mathbf{x},\mathbf{Y}=\mathbf{y})$,
$P(\mathbf{y}|\mathbf{x})\dfn\P(\mathbf{Y}=\mathbf{y}|\mathbf{X}=\mathbf{x})$,
and $P(\mathbf{x}|\mathbf{y})\dfn\P(\mathbf{X}=\mathbf{x}|\mathbf{Y}=\mathbf{y})$.
Let $R$ be the coding rate in nats per channel use, and let ${\cal C}=\{\mathbf{x}_{1},\mathbf{x}_{2}\ldots,\mathbf{x}_{M}\}$,
$\mathbf{x}_{m}\in{\cal X}^{n}$, $m=1,\ldots,M$, $M=e^{nR}$ denote
the codebook.

An erasure/list decoder $\phi$ is a mapping from the space of output
vectors ${\cal Y}^{n}$ to the set of subsets of $\{1,\ldots,M\}$
(i.e., $\phi:{\cal Y}^{n}\to{\cal P}(\{1,\ldots,M\})$, where the
latter is the power set of $\{1,\ldots,M\}$). Alternatively, an erasure/list
decoder $\phi$ is uniquely defined by a set of $M+1$ decoding regions
$\{{\cal R}_{m}\}_{m=0}^{M}$ such that ${\cal R}_{m}\subseteq{\cal Y}^{n}$
and ${\cal R}_{0}={\cal Y}^{n}\backslash\bigcup_{m=1}^{M}{\cal R}_{m}$.
Given an output vector $\mathbf{y}$, the $m$th codeword belongs
to the list if $\mathbf{y}\in{\cal R}_{m}$, and if $\mathbf{y}\in{\cal R}_{0}$
an erasure is declared.

The average error probability of a list decoder $\phi$ and a codebook
${\cal C}$ is the probability that the actual transmitted codeword
does not belong to the list, 
\[
P_{e}({\cal C},\phi)\dfn\frac{1}{M}\sum_{m=1}^{M}\sum_{\mathbf{y}\in\overline{{\cal R}}_{m}}P(\mathbf{y}|\mathbf{x}_{m}).
\]
The\emph{ }average number of erroneous codewords on the list is defined
as 
\begin{align*}
L({\cal C},\phi) & \dfn\sum_{\mathbf{y}\in{\cal Y}^{n}}\frac{1}{M}\sum_{m=1}^{M}P(\mathbf{y}|\mathbf{x}_{m})\sum_{l\neq m}\I(\mathbf{y}\in{\cal R}_{l})\\
 & =\frac{1}{M}\sum_{l=1}^{M}\sum_{\mathbf{y}\in{\cal R}_{l}}\sum_{m\neq l}P(\mathbf{y}|\mathbf{x}_{m}).
\end{align*}
If the decoder never outputs more than one codeword, then it is termed
a \emph{pure erasure} decoder. In this case, $P_{e}({\cal C},\phi)$
designates the \emph{total error probability} (which is the sum of
the \emph{erasure probability }and \emph{undetected error probability}),
and $L({\cal C},\phi)$ designates the \emph{undetected error probability.
}In what follows, we shall describe our results with the more general
erasure/list terms, but unless otherwise stated, the results are also
valid for pure erasure decoders, with the above interpretation. 

In \cite{Forney68}, Forney has generalized the Neyman-Pearson lemma,
and obtained a class of decoders $\Phi^{*}\dfn\{\phi_{T},T\in\reals\}$
which achieves the optimal trade-off between $P_{e}({\cal C},\phi)$
and $L({\cal C},\phi)$. The decoding regions for a decoder $\phi_{T}\in\Phi^{*}$
are given by 
\begin{equation}
{\cal R}_{m}^{*}\dfn\left\{ \mathbf{y}:\frac{P(\mathbf{y}|\mathbf{x}_{m})}{\sum_{l\neq m}P(\mathbf{y}|\mathbf{x}_{l})}\geq e^{nT}\right\} ,\,\,\,\,\,1\leq m\leq M.\label{eq: optimal decoding regions Forney}
\end{equation}
The parameter $T$ is called the \emph{threshold}, and it controls
the trade-off between $P_{e}({\cal C},\phi)$ and $L({\cal C},\phi)$.
As $T$ increases, the list size typically becomes smaller, but in
exchange, the error probability increases. To obtain a pure erasure
decoder, the threshold $T$ should be chosen non-negative. For $T<0$,
an erasure/list decoder is obtained.

As a figure of merit for a decoder, we will consider its random coding
exponents, over the ensemble of fixed composition codes with input
distribution $P_{X}$. For this ensemble, the $M=e^{nR}$ codewords
of ${\cal C}=\{\mathbf{x}_{1},\ldots,\mathbf{x}_{M}\}$ are selected
independently at random under the uniform distribution across the
type class ${\cal T}_{P_{X}}$. We denote the averaging over this
ensemble of random codebooks by overlines. The random coding error
exponent is defined as
\[
E_{e}(R,\phi)\dfn\liminf_{n\to\infty}-\frac{1}{n}\log\overline{P_{e}({\cal C},\phi)}
\]
and the random coding list size (negative) exponent is defined as
\[
E_{l}(R,\phi)\dfn\liminf_{n\to\infty}-\frac{1}{n}\log\overline{L({\cal C},\phi)}.
\]
Notice that $E_{e}(R,\phi)\geq0$ but $E_{l}(R,\phi)$ may also be
negative, which means that the random coding list size may \emph{increase}
exponentially. 

In \cite{exact_erasure}, the exact random coding exponents for a
decoder $\phi_{T}^{*}\in\Phi^{*}$, which we denote by $E_{e}^{*}(R,T)$
and $E_{l}^{*}(R,T)$, were found%
\footnote{See Appendix \ref{sec:Justification-of-the} for a justification of
the applicability of the results in \cite{exact_erasure} also for
the case $T<0$.%
}. Defining 
\begin{equation}
E_{a}(R,T)\dfn\min_{\tilde{Q}}\min_{Q:\: Q_{Y}=\tilde{Q}_{Y},f(Q)+T\geq f(\tilde{Q}),I(Q)\geq R}\left\{ D(\tilde{Q}_{Y|X}||W|P_{X})+I(Q)\right\} -R,\label{eq: error exponent optimal Ea}
\end{equation}
and
\begin{equation}
E_{b}(R,T)\dfn\min_{\tilde{Q}\in{\cal L}}D(\tilde{Q}_{Y|X}||W|P_{X}),\label{eq: error exponent optimal Eb}
\end{equation}
where 
\[
{\cal L}\dfn\left\{ \tilde{Q}:f(\tilde{Q})\leq R+T+\max_{Q:\: Q_{Y}=\tilde{Q}_{Y},I(Q)\leq R}\left[f(Q)-I(Q)\right]\right\} ,
\]
we have that 
\begin{equation}
E_{e}^{*}(R,T)=\min\{E_{a}(R,T),E_{b}(R,T)\}\label{eq: error exponent optimal}
\end{equation}
and \cite[Lemma 1]{universal_erasure_exact} 
\begin{equation}
E_{l}^{*}(R,T)=E_{e}^{*}(R,T)+T.\label{eq:list exponent optimal}
\end{equation}

In the rest of the paper, we will consider simplified erasure/list
decoding which determine if a codeword belongs to the list, by checking
if the likelihood of the codeword is larger than some threshold function.
To motivate this approach, we first inspect the optimal decoder \eqref{eq: optimal decoding regions Forney}
in more detail. In its current form, the optimal decoder is essentially
infeasible to implement. The reason for this is that the computation
of $\sum_{l\neq m}P(\mathbf{y}|\mathbf{x}_{l})$ is usually intractable,
as it is a sum of exponentially many likelihood terms, where each
likelihood term is also exponentially small. This is in sharp contrast
to ordinary decoders, based on comparison of single likelihood terms
which can be carried out in the logarithmic scale, rendering them
numerically feasible. However, the optimal decoder can be simplified
without asymptotic loss in exponents. To see this, we need first some
notation: for a given joint type $Q$ and output vector $\mathbf{y}$,
let $N_{m}(Q|\mathbf{y})$ be the number of codewords other than $\mathbf{x}_{m}$
in ${\cal C}$, whose joint empirical distribution with $\mathbf{y}$
is $Q$. Now, define the decision regions
\begin{equation}
\tilde{{\cal R}}_{m}\dfn\left\{ \mathbf{y}:e^{nf(\hat{Q}_{\mathbf{x}_{m}\mathbf{y}})}\geq e^{nT}\max_{Q}N_{m}(Q|\mathbf{y})\cdot e^{nf(Q)}\right\} ,\,\,\,\,\,1\leq m\leq M,\label{eq: simplified optimal decision regions}
\end{equation}
where $f(Q)$ is the normalized log-likelihood ratio, namely 
\[
f(Q)\dfn\sum_{x\in{\cal X},y\in{\cal Y}}Q(x,y)\log W(y|x).
\]
We have the following observation.
\begin{prop}
\label{prop: simplified optimal decoder}The random coding error-
and list size exponents of the decoder defined by $\{\tilde{{\cal R}}_{m}\}_{m=1}^{M}$
are the same as the random coding error- and list size exponents of
the optimal decoder (defined by $\{{\cal R}_{m}^{*}\}_{m=1}^{M}$).
\end{prop}
We will refer to the right-hand side of \eqref{eq: simplified optimal decision regions}
as the \emph{threshold} of the decoder. The focus of this paper, is
the performance of erasure/list decoders for which the threshold function
is not necessarily the optimal threshold dictated by \eqref{eq: simplified optimal decision regions}.
The motivation for such an erasure/list decoder is that the threshold
function employed may be simpler to compute. Observe, that the threshold
of \eqref{eq: simplified optimal decision regions} only depends on
the joint type of each competing codeword ($l\neq m$) with the output
vector. Therefore, we propose to consider the class of decoders $\Psi\dfn\{\phi_{h},h:{\cal S}({\cal X}\times{\cal Y})\to\reals\}$,
defined by a continuous function $h(\cdot)$ in some compact domain
${\cal G}\subseteq{\cal S}({\cal X}\times{\cal Y})$, and infinite
elsewhere%
\footnote{The set ${\cal G}$ will be a strict subset of ${\cal S}({\cal X}\times{\cal Y})$
for optimal threshold functions, see Section \ref{sec:Optimal-Choice-of}.%
}. The decoding regions of $\phi_{h}\in\Psi$ are given by
\begin{equation}
{\cal R}_{m}=\left\{ \mathbf{y}:e^{nf(\hat{Q}_{\mathbf{x}_{m}\mathbf{y}})}\geq e^{n\cdot\max_{l\neq m}h(\hat{Q}_{\mathbf{x}_{l}\mathbf{y}})}\right\} ,\,\,\,\,\,1\leq m\leq M.\label{eq: exponential threhsold decision regions}
\end{equation}
As discussed above, for this decoding rule, computations may be carried
in the logarithmic domain as
\[
{\cal R}_{m}=\left\{ \mathbf{y}:f(\hat{Q}_{\mathbf{x}_{m}\mathbf{y}})\geq\max_{l\neq m}h(\hat{Q}_{\mathbf{x}_{l}\mathbf{y}})\right\} .
\]
With a slight abuse of notation, we shall denote the random coding
exponents of $\phi_{h}\in\Psi$ by $E_{e}(R,h)$ and $E_{l}(R,h)$.
Note that in \eqref{eq: exponential threhsold decision regions},
the threshold function is assumed to be the same for all codebooks
in the ensemble (for specific rate $R$). In contrast, for the asymptotically
optimal decoder \eqref{eq: simplified optimal decision regions},
the threshold is tuned to the specific codebook employed. This modification
reduces complexity (though only slightly) and may, in general, degrade
the exponents. Nonetheless, for types which satisfy $I(Q)\leq R$,
it is known that when the $m$th codeword is sent, $N_{m}(Q|\mathbf{y})$
concentrates double-exponentially fast around its asymptotic expected
value of $e^{n(R-I(Q))}$ \cite[Section 6.3]{Neri_statistical_physics}.
This hints to the possibility that no loss in exponents is incurred
by the restriction of a single threshold function used for all codebooks
in the ensemble. Consequently, for a properly chosen $h(\cdot)$,
it is possible that $\phi_{h}$ achieves the optimal exponents. 

Next, we propose two subclasses of $\Psi$. The first subclass is
$\Lambda_{1}\dfn\{\phi_{g},g:{\cal S}({\cal Y})\to\reals\}$ where
$g(\cdot)$ is a continuous function in some compact domain ${\cal G}\subseteq{\cal S}({\cal Y})$,
and infinite elsewhere. A decoder $\phi_{g}\in\Lambda_{1}$ has the
following decoding regions 
\[
{\cal R}_{m}=\left\{ \mathbf{y}:f(\hat{Q}_{\mathbf{x}_{m}\mathbf{y}})\geq g(\hat{Q}_{\mathbf{y}})\right\} ,\,\,\,\,\,1\leq m\leq M.
\]
With a slight abuse of notation, we shall denote the random coding
exponents of $\phi_{g}\in\Lambda_{1}$ by $E_{e}(R,g)$ and $E_{l}(R,g)$.
Here, the \emph{threshold }function $g(Q_{Y})$ does not depend on
the codebook, but it may depend on $R$. In essence, a decoder in
$\Lambda_{1}$ approximates $e^{nT}N_{m}(Q|\mathbf{y})\cdot e^{nf(Q)}$
by a function that depends on $\mathbf{y}$ only, but not on ${\cal C}$.
The second subclass is $\Lambda_{2}\dfn\{\phi_{T},T\in\reals\}$,
where $T$ is a threshold parameter. A decoder $\phi_{T}\in\Lambda_{2}$
has the following decoding regions, 
\[
{\cal R}_{m}=\left\{ \mathbf{y}:f(\hat{Q}_{\mathbf{x}_{m}\mathbf{y}})\geq T+\max_{l\neq m}f(\hat{Q}_{\mathbf{x}_{l}\mathbf{y}})\right\} ,\,\,\,\,\,1\leq m\leq M.
\]
With a slight abuse of notation, we shall denote the random coding
exponents of $\phi_{T}\in\Lambda_{2}$ by $E_{e}(R,T)$ and $E_{l}(R,T)$.
Observe that for $T<0$, the list size of this decoder is at least
$1$, since the codeword with maximum likelihood is always on the
list. In essence, this decoder approximates 
\[
N_{m}(Q|\mathbf{y})\cdot e^{nf(Q)}\approx\max_{N_{m}(Q|\mathbf{y})\geq1}N_{m}(Q|\mathbf{y})\cdot e^{nf(Q)}\approx\max_{l\neq m}P(\mathbf{y}|\mathbf{x}_{l}),
\]
and so, the threshold, in this case, is the second largest likelihood.
This approximation was proposed by Forney \cite[eq. (11a)]{Forney68},
but was not analyzed rigorously before.

In this paper, we provide \emph{exact} single-letter expressions for
the error- and list size exponents for the class $\Psi$, and obtain,
as corollaries, the exponents of the previously defined subclasses
$\Lambda_{1}$ and $\Lambda_{2}$. Then, for each of the classes of
decoders, we will derive the \emph{optimal }threshold function, in
the sense that for a given requirement on the value of $E_{e}(R,\phi)$,
they provide the largest $E_{l}(R,\phi)$ within the given class.
Finally, we discuss the regimes for which the simplified decoders
are close to be asymptotically optimal. For the defined ensemble of
random codes, the subclasses $\Lambda_{1}$ and $\Lambda_{2}$ represent
two extremes of approximation to the threshold. For low rates, and
a typical codebook in the ensemble, the threshold will be dominated
by a single codeword,%
\footnote{At least, a sub-exponential number of codewords.%
} which has the maximum likelihood, besides the candidate codeword%
\footnote{This could be either the codeword which was actually sent, or an erroneous
codeword.%
}, and the random coding exponents of $\phi_{T}\in\Lambda_{2}$ will
be optimal. On the other hand, for high rates, the threshold will
tend to concentrate around a deterministic function $g(\hat{Q}_{\mathbf{y}})$.
Using the function $g(\cdot)$ as a threshold function will be accurate
for a typical codebook, and so, in this case too, the random coding
exponents of $\phi_{g}\in\Lambda_{1}$ will tend to the optimal exponents,
for high rates. 

Before we continue, we emphasize that our decoders do not assume any
specific structure of the codebook (as we have assumed the fixed composition
random coding ensemble), and so the simplified decoders do no immediately
lead to practical implementation. Nonetheless, as common, the random
coding analysis serves as an achievable benchmark for possible exponents.

\section{Exponents of Threshold Decoders\label{sec:Exponents-of-Threshold}}

In this section, we derive the error- and list size exponents for
the class $\Psi$, and then obtain, as special cases, the exponents
for the subclasses $\Lambda_{1}$ and $\Lambda_{2}$\emph{. }As we
have assumed an ensemble of fixed composition codebooks with input
distribution $P_{X}$, a joint type $Q$ of $(\mathbf{x}_{m},\mathbf{y})$
will always have the form $P_{X}\times Q_{Y|X}$. For brevity, this
is not explicitly mentioned henceforth. We begin with our main theorem,
which provides the random coding exponents for $\phi_{h}\in\Psi$.
We use the following definitions. For a given $h(\cdot)$, let 

\begin{equation}
\mathbf{V}(Q_{Y},R)\dfn\max_{Q':\: Q'_{Y}=Q_{Y},I(Q')\leq R}h(Q'),\label{eq: V_bold definition}
\end{equation}
and 
\begin{equation}
{\cal D}(\tilde{Q})\dfn\left\{ Q:Q=\tilde{Q}_{Y},f(Q)\geq\max\left\{ \mathbf{V}(Q_{Y},R),h(\tilde{Q})\right\} \right\} .\label{eq: cal=00007BD=00007D definition}
\end{equation}

\begin{thm}
\label{thm:exp threshold exponents}For a decoder $\phi_{h}\in\Psi$,
the random coding error exponent, with respect to (w.r.t) the ensemble
of fixed composition codebooks $P_{X}$, is given by
\begin{equation}
E_{e}(R,h)=\min_{\tilde{Q}}\min_{Q:\: Q_{Y}=\tilde{Q}_{Y},h(Q)\geq f(\tilde{Q})}\left\{ D(\tilde{Q}_{Y|X}||W|P_{X})+\left[I(Q)-R\right]_{+}\right\} ,\label{eq: error exponent exponential threshold}
\end{equation}
and the list size exponent is given by 
\begin{equation}
E_{l}(R,h)=\min_{\tilde{Q}}\min_{Q\in{\cal D}(\tilde{Q})}\left\{ D(\tilde{Q}_{Y|X}||W|P_{X})+I(Q)-R\right\} .\label{eq: list exponent exponential threshold}
\end{equation}

\end{thm}
Next, we provide the exponents for the subclasses $\Lambda_{1}$ and
$\Lambda_{2}$. These exponents are obtained using the general result
of Theorem \ref{thm:exp threshold exponents}, with appropriate substitutions
and manipulations. 
\begin{cor}
\label{cor:output exponents}For a decoder $\phi_{g}\in\Lambda_{1}$,
the random coding error exponent, w.r.t. the ensemble of fixed composition
codebooks $P_{X}$, is given by 
\begin{equation}
E_{e}(R,g)=\min_{Q:\: g(Q_{Y})\geq f(Q)}D(Q_{Y|X}||W|P_{X}),\label{eq: error exponent output}
\end{equation}
and the list size exponent is given by

\begin{equation}
E_{l}(R,g)=\min_{\tilde{Q}}\min_{Q:\: Q_{Y}=\tilde{Q}_{Y},g(Q_{Y})\leq f(Q)}\left\{ D(\tilde{Q}_{Y|X}||W|P_{X})+I(Q)-R\right\} .\label{eq: list exponent output}
\end{equation}

\end{cor}
Note that, for a given $g(\cdot)$, the error exponent $E_{e}(R,g)$
does not depend on $R$, because the decision whether to include a
codeword in the list is based only the codeword sent and the output
vector received, not on other codewords. Next, for $T<0$, we define
\[
\underline{\mathbf{V}}(Q_{Y},R)\dfn\max_{Q':\: Q'_{Y}=Q_{Y},I(Q')\leq R,}f(Q'),
\]
and 
\[
\underline{{\cal D}}(\tilde{Q},R,T)\dfn\left\{ Q:Q_{Y}=\tilde{Q}_{Y},f(Q)\geq T+\max\left\{ \underline{\mathbf{V}}(Q_{Y},R),f(\tilde{Q})\right\} \right\} .
\]

\begin{cor}
\label{cor:ML exponents}For a decoder $\phi_{T}\in\Lambda_{2}$,
the random coding error exponent, w.r.t. the ensemble of fixed composition
codebooks $P_{X}$, is given by 
\begin{equation}
E_{e}(R,T)=\min_{\tilde{Q}}\min_{Q:\: Q_{Y}=\tilde{Q}_{Y},f(Q)+T\geq f(\tilde{Q})}\left\{ D(\tilde{Q}_{Y|X}||W|P_{X})+\left[I(Q)-R\right]_{+}\right\} ,\label{eq: error exponent ML}
\end{equation}
and the list size exponent is given by

\begin{equation}
E_{l}(R,T)=\min_{\tilde{Q}}\min_{Q\in\underline{{\cal D}}(\tilde{Q},R,T)}\left\{ D(\tilde{Q}_{Y|X}||W|P_{X})+I(Q)-R\right\} .\label{eq: list exponent ML}
\end{equation}
\end{cor}
\begin{rem}
\label{rem:mismatch}Assume that for some channel $V\neq W$, $f(Q)$
is replaced by 
\[
\tilde{f}(Q)\dfn\sum_{x\in{\cal X},y\in{\cal Y}}Q(x,y)\log V(y|x)
\]
in the optimal exponents \eqref{eq: error exponent optimal Ea}--\eqref{eq:list exponent optimal},
as well as in Theorem \ref{thm:exp threshold exponents} and its corollaries
\ref{cor:output exponents} and \ref{cor:ML exponents}. This yields
random coding exponents associated with \emph{mismatched decoding}.
\end{rem}

\section{Optimal Threshold Functions\label{sec:Optimal-Choice-of}}

In the previous section, we have derived the exact exponents for decoders
of the classes $\Psi$, $\Lambda_{1}$ and $\Lambda_{2}$, where these
exponents depend on the actual decoder in the class via the threshold
functions $h(\cdot)$, $g(\cdot)$, and the $T$, respectively. In
some scenarios, there is a freedom to choose any decoder within a
given class. Clearly, just as for the optimal class $\Phi^{*}$, a
trade-off exists between the error- and list size exponents, which
is controlled by the choice of threshold function or parameter. In
this section, we assume a given rate $R$, and a target error exponent
$\EE$. Under this requirement, we find the threshold function or
parameter for a given class of decoders, which yields the maximal
list size exponent. In addition, we find expressions for the resulting
maximal list size exponent. Obviously, the resulting list size exponent
cannot exceed $E_{l}^{*}(R,T^{*})$, where $T^{*}$ satisfies $E_{e}^{*}(R,T^{*})=\EE$,
and the difference between these two exponents is a measure for the
sub-optimality of the \emph{class} of decoders. We first define optimal
threshold functions. 
\begin{defn}
A threshold function $h^{*}(Q,R,\EE)$ is said to be \emph{optimal}
for the class $\Psi$, if the corresponding decoder $\phi_{h^{*}(Q,R,\EE)}\in\Psi$
achieves $E_{e}(R,h)\geq\EE$ and for any other decoder $\phi_{h}\in\Psi$
with $E_{e}(R,h)\geq\EE$, we have $E_{l}(R,h)\leq E_{l}(R,h^{*})$. 
\end{defn}
The optimal threshold function $g^{*}(Q_{Y},\EE)$ for the class $\Lambda_{1}$,
and the optimal parameter $T^{*}(R,\EE)$ for the class $\Lambda_{2}$,
are defined analogously. The resulting list size exponent for $h^{*}(Q,R,\EE)$
will be denoted by $E_{l}^{*}(\Psi,R,\EE)$, and $E_{l}^{*}(\Lambda_{1},R,\EE)$,
$E_{l}^{*}(\Lambda_{2},R,\EE)$, will denote the analogue exponents
for $g^{*}(Q_{Y},\EE)$ and $T^{*}(R,\EE)$. In this section, we will
find it more convenient to begin with the simple class $\Lambda_{1}$,
as $h^{*}(Q,R,\EE)$ is conveniently represented by $g^{*}(Q_{Y},\EE)$.
We then conclude with $T^{*}(R,\EE)$ for the class $\Lambda_{2}$,
which is conveniently represented by $h^{*}(Q,R,\EE)$. 
\begin{thm}
\label{thm: optimal threshold output}The optimal threshold function
$g^{*}(Q_{Y},\EE)$ for the class $\Lambda_{1}$ is
\begin{equation}
g^{*}(Q_{Y},\EE)=\min_{Q':\: Q'{}_{Y}=Q_{Y},D(Q'{}_{Y|X}||W|P_{X})\leq\EE}f(Q'),\label{eq: g_star definition}
\end{equation}
and the resulting list size exponent is 
\[
E_{l}^{*}(\Lambda_{1},R,\EE)=\min_{\tilde{Q}}\min_{Q:\: Q_{Y}=\tilde{Q}_{Y},g^{*}(Q_{Y},\EE)\leq f(Q)}\left\{ D(\tilde{Q}_{Y|X}||W|P_{X})+I(Q)-R\right\} .
\]

\end{thm}
Note that for $\EE<0$, the minimization problem in \eqref{eq: g_star definition}
is infeasible and so $g^{*}(Q_{Y},\EE)=\infty$. While $g^{*}(Q_{Y},\EE)$
does not depend directly on the rate $R$, the required error exponent
$\EE$ will usually be chosen as a function of the rate $R$, and
thus $g^{*}(Q_{Y},\EE)$ will depend indirectly on $R$. The next
lemma states a few simple properties of $g^{*}(Q_{Y},\EE)$.
\begin{lem}
\label{lem: Optimal output threshold properties}The optimal threshold
function $g^{*}(Q_{Y},\EE)$ has the following properties:
\begin{enumerate}
\item It is a non-increasing function of $\EE$. \label{enu: output optimal non increasing}
\item If $D(Q_{Y|X}||W|P_{X})\leq\EE$ then $f(Q)\geq g^{*}(Q_{Y},\EE)$.\label{enu: output optimal divergence}
\item It is a convex function of $Q_{Y}$.\label{enu: output optimal convex of Qy}
\end{enumerate}
\end{lem}
Next, we provide the optimal list size exponent for the class $\Psi$.
We define
\begin{align}
{\cal D}^{*}(\tilde{Q},R,\EE) & =\left\{ Q:Q_{Y}=\tilde{Q}_{Y},f(Q)>g^{*}(Q_{Y},\EE-[I(\tilde{Q})-R]_{+})\right\} .\label{eq: D* for optimal exponential threshold}
\end{align}

\begin{thm}
\label{thm: optimal threshold unified}The optimal threshold function
$h^{*}(Q,R,\EE)$ for the class $\Psi$ is
\begin{equation}
h^{*}(Q,R,\EE)=g^{*}(Q_{Y},\EE-\left[I(Q)-R\right]_{+})\label{eq: optimal threhsold exponential}
\end{equation}
and the resulting list size exponent is
\begin{equation}
E_{l}^{*}(\Psi,R,\EE)=\min_{\tilde{Q}}\min_{Q\in{\cal D}^{*}(\tilde{Q},R,\EE)}\left\{ D(\tilde{Q}_{Y|X}||W|P_{X})+I(Q)-R\right\} .\label{eq: list exponent optimal exponential threshold proof}
\end{equation}

\end{thm}
It is easily shown, using Lemma \ref{lem: Optimal output threshold properties}
(property \ref{enu: output optimal divergence}), that in the domain
where $h^{*}(Q,R,\EE)<\infty$, it is a continuous function of the
joint type $Q$. 
\begin{rem}
Following Remark \ref{rem:mismatch}, consider the scenario for which
the channel $W$ is not known exactly, but only known to belong to
a given class of channels ${\cal W}$. For example, suppose we are
given a nominal channel $V$, and it is known that $W$ is not far
from $V$, where the distance is measured in the ${\cal L}_{1}$ norm.
For a moment, let us denote the optimal threshold, as $g_{W}^{*}(Q_{Y},\EE)$,
i.e. with explicit dependency in the channel $W$. Then, choosing
a threshold function 
\[
g_{{\cal W}}^{*}(Q_{Y},\EE)\triangleq\min_{W\in{\cal W}}g_{W}^{*}(Q_{Y},\EE)
\]
guarantees that $E_{e}(R,g_{W}^{*})\geq\EE$ uniformly over $W\in{\cal W}$.
The resulting list size exponent 
\begin{align*}
E_{l}(R,g) & =\min_{\tilde{Q}}\min_{Q:\: Q_{Y}=\tilde{Q}_{Y},\min_{W\in{\cal W}}g_{W}^{*}(Q_{Y},\EE)\leq f(Q)}\left\{ D(\tilde{Q}_{Y|X}||W|P_{X})+I(Q)-R\right\} \\
 & =\min_{W\in{\cal W}}\min_{\tilde{Q}}\min_{Q:\: Q_{Y}=\tilde{Q}_{Y},g_{W}^{*}(Q_{Y},\EE)\leq f(Q)}\left\{ D(\tilde{Q}_{Y|X}||W|P_{X})+I(Q)-R\right\} \\
 & =\min_{W\in{\cal W}}E_{l}^{*}(W,\Lambda_{1},R,\EE)
\end{align*}
where here, $E_{l}^{*}(W,\Lambda_{1},R,\EE)$ denotes the optimal
list size exponent of Theorem \ref{thm: optimal threshold output},
with explicit dependency in $W$. Analogue results hold also for the
class $\Psi$. See \cite{Moulin_universal_erasure,universal_minimax_erasure,universal_erasure_exact}
for related ideas. 
\end{rem}
We conclude this section with the optimal $T$ for the class $\Lambda_{2}$. 
\begin{thm}
\textbf{\label{thm: optimal threshold ML}}The optimal threshold parameter
$T^{*}(R,\EE)$ for the class $\Lambda_{2}$ is
\begin{equation}
T^{*}(R,\EE)=\min_{Q}\left\{ h^{*}(Q,R,\EE)-f(Q)\right\} \label{eq: optimal threhsold ML}
\end{equation}
and the resulting list size exponent is 
\begin{equation}
E_{l}^{*}(\Lambda_{2},R,\EE)=\min_{\tilde{Q}}\min_{Q\in\underline{{\cal D}}(\tilde{Q},R,T^{*})}\left\{ D(\tilde{Q}_{Y|X}||W|P_{X})+I(Q)-R\right\} .\label{eq: list size optimal exponent ML threshold}
\end{equation}

\end{thm}

\section{Optimality of Simplified Decoders for Low and High Rates\label{sec:Optimality-of-Simplified}}

In this section, we discuss the optimality of the two subclasses $\Lambda_{1}$
and $\Lambda_{2}$ in certain regimes of the rate. For $\Lambda_{1}$,
we show that for rates \emph{above} some critical rate, the optimal
decoder from the class $\Lambda_{1}$ has the same exponents as the
optimal decoder from the class $\Psi$. In turn, as we have discussed
in Subsection \ref{sub:System-Model-and}, the optimal decoder from
the class $\Psi$ has exponents which are close, and sometimes even
equal, to the optimal exponents of $\phi_{T}^{*}\in\Phi^{*}$. For
$\Lambda_{2}$, we show that assuming $T>0$, for rates \emph{below}
some critical rate, the exponents of the decoder $\phi_{T}\in\Lambda_{2}$
are the same exponents as $\phi_{T}^{*}$ (for the same value of $T$).
From continuity arguments, approximate optimality will be obtained
for $T<0$ with small $|T|$. Moreover, for $T<0$, or for rates above
the critical rate, the exponents of the decoder from $\Lambda_{2}$
will improve if we choose the optimal $T^{*}$, according to Theorem
\ref{thm: optimal threshold ML}.

We begin with $\Lambda_{1}$. Let the optimal solution of $E_{l}^{*}(\Lambda_{1},R,\EE)$
be $(\tilde{Q}^{*},Q^{*})$, and let $\overline{R}_{\s[cr]}(\EE)\dfn I(\tilde{Q}^{*})$. 
\begin{prop}
\label{prop: optimality of output threshold}For $R\geq\overline{R}_{\s[cr]}(\EE)$
\[
E_{l}^{*}(\Lambda_{1},R,\EE)=E_{l}^{*}(\Psi,R,\EE).
\]
Namely, a threshold which only depends on the output vector is sufficient
to obtain the best exponents of the class $\Psi$.
\end{prop}
Next, we demonstrate the optimality of the class $\Lambda_{2}$ in
case that $R$ is not too large, and $T\geq0$. Consider a decoder
$\phi_{T}\in\Lambda_{2}$, and an optimal decoder $\phi_{T}^{*}\in\Phi^{*}$,
with the same parameter $T\geq0$. Let now $(\tilde{Q}^{*},Q^{*})$
be the optimal solution of 
\[
\min_{\tilde{Q}}\min_{Q:\: Q_{Y}=\tilde{Q}_{Y},f(Q)+T\geq f(\tilde{Q})}\left\{ D(\tilde{Q}_{Y|X}||W|P_{X})+I(Q)\right\} ,
\]
and let $\underline{R}_{\s[cr]}(T)\dfn I(\tilde{Q}^{*})$. 
\begin{prop}
\label{prop: optimality of ML threshold}For $T\geq0$ and $R\leq\underline{R}_{\s[cr]}(T)$
both 
\[
E_{e}(R,T)=E_{e}^{*}(R,T),
\]
and
\[
E_{l}(R,T)=E_{l}^{*}(R,T).
\]
Namely, $\phi_{T}\in\Lambda_{2}$ and the optimal $\phi_{T}^{*}\in\Phi^{*}$
have the same exponents.
\end{prop}

\section{Numerical Examples\label{sec:Numerical-Example}}

We demonstrate the results for binary channels ($|{\cal X}|=|{\cal Y}|=2$),
assuming a threshold $T=\pm0.05$. First, for any rate $0\leq R\leq I(P_{X}\times W)$,
the optimal exponents $E_{e}^{*}(R,T)$ and $E_{l}^{*}(R,T)$ were
computed using \eqref{eq: error exponent optimal}, and \eqref{eq:list exponent optimal}.
Second, for any rate $0\leq R\leq I(P_{X}\times W)$, the target error
exponent was set to $\EE=E_{e}^{*}(R,T)$, and the maximal list size
exponent was found for the class $\Psi$, as well as its subclasses
$\Lambda_{1}$ and $\Lambda_{2}$. 

\begin{figure}
\centering{}\includegraphics[scale=0.5]{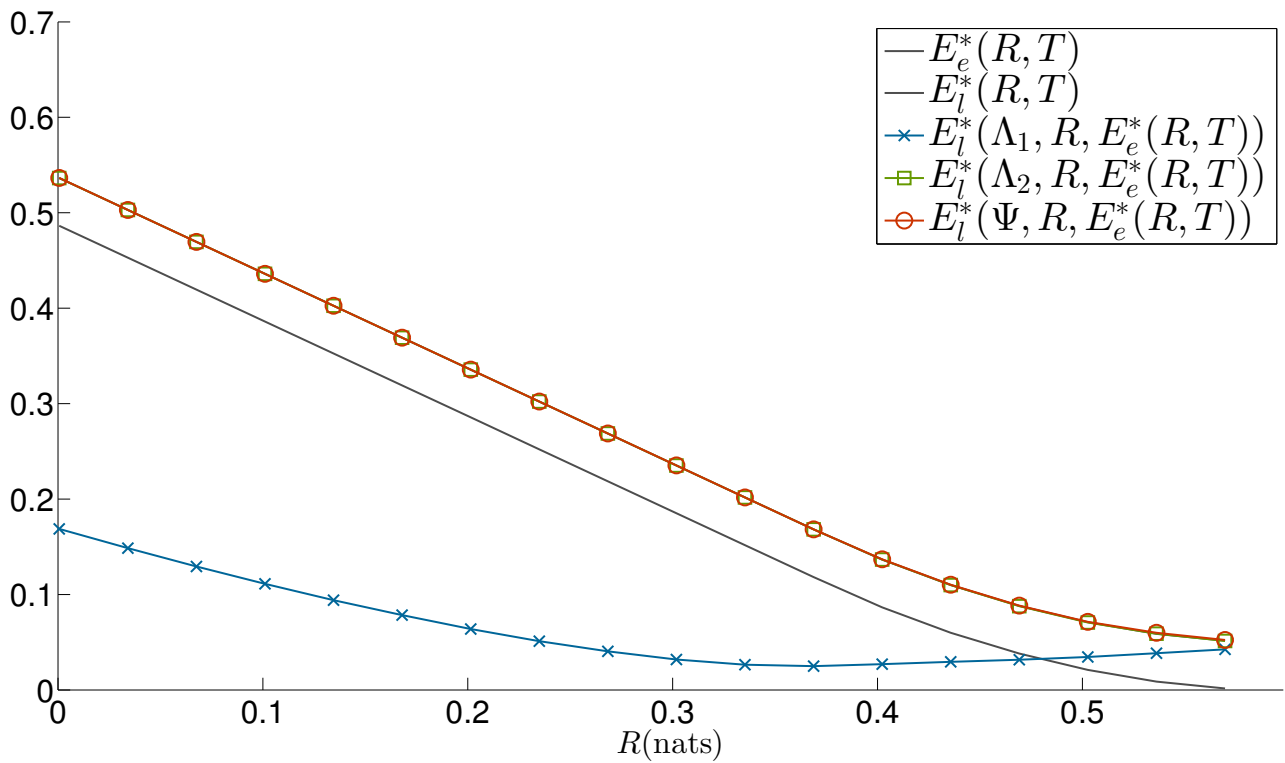}\protect\caption{Graphs of random coding exponents for $W_{1}$ and $T=0.05$.\label{fig:BSC-example}}
\end{figure}

For $T=0.05$, Figure \ref{fig:BSC-example} shows the random coding
exponents for a binary symmetric channel $W_{1}(0|1)=W_{1}(1|0)=0.01$.
In this example, the optimal decoder from $\Lambda_{2}$ has the same
exponents as $\phi_{T}^{*}$ for the entire range of rates. It is
interesting to note that for the optimal decoder from $\Lambda_{1}$,
the maximal list size is not a monotonic decreasing function of the
rate (but naturally always smaller than the optimal list size exponent,
for the given error exponent). This is due to the two contradicting
effects of the rate on the optimal list size exponent of the class
$\Lambda_{1}$: As the rate increases, the class $\Lambda_{1}$ has
improved performance on one hand, but the error exponent requirement
is decreasing (as we have assumed a fixed $T$) on the other hand. 

\begin{figure}
\centering{}\includegraphics[scale=0.5]{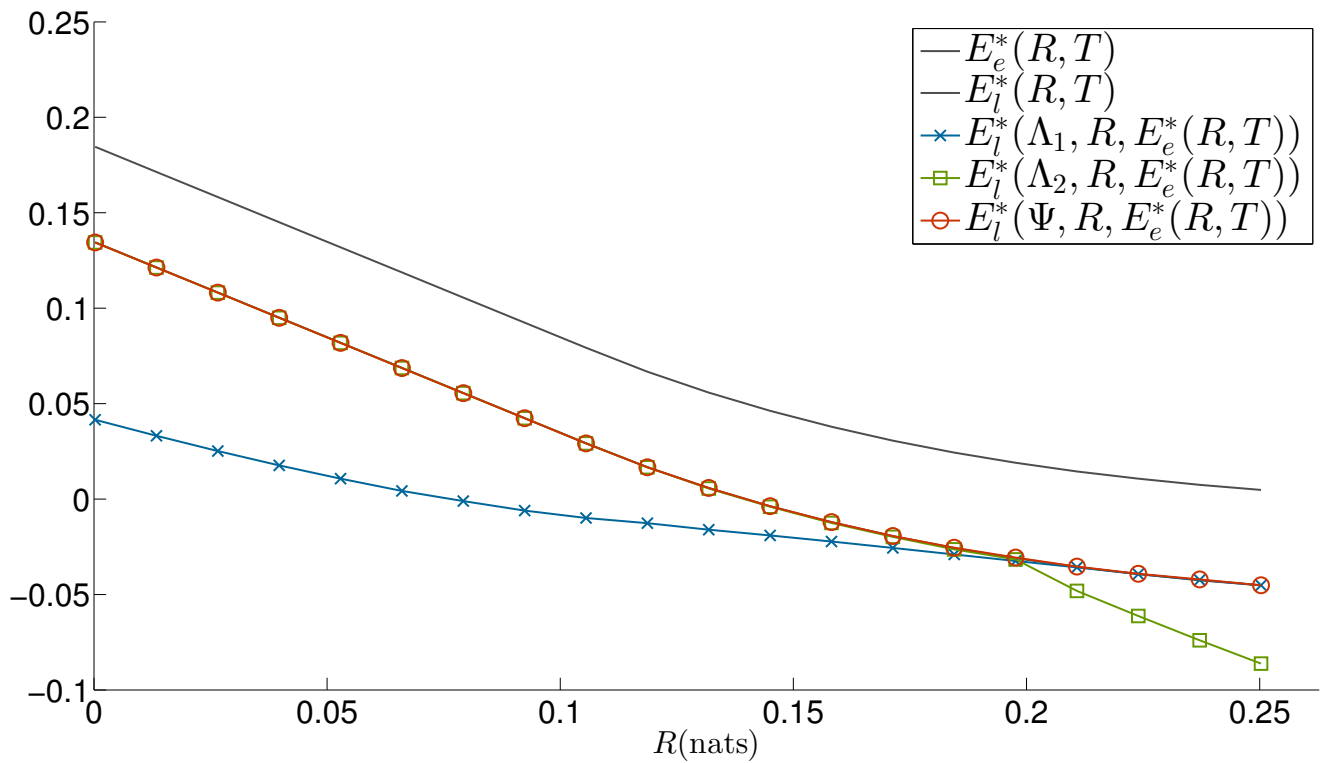}\protect\caption{Graphs of random coding exponents for $W_{2}$ and $T=-0.05$.\label{fig:BNSC-example}}
\end{figure}

Next, for $T=-0.05$, Figure \ref{fig:BNSC-example}, shows the exponents
for a binary asymmetric channel $W_{2}(0|1)=0.01,W_{2}(1|0)=0.4$.
For high rates, the optimal decoder from $\Lambda_{1}$ approaches
optimal performance, but is rather poor for low rates. For these low,
and even intermediate, rates, the optimal decoder from $\Lambda_{2}$
has performance close to optimal. It was also found empirically, that
the rate for which the maximum likelihood threshold decoder moves
away from optimal performance is larger as the channel is more symmetric.
Thus, the performance of the simplified decoders for $T=-0.05$ is
even better for the previously defined symmetric channel. 

In both examples, it can be observed that for the \emph{entire} range
of rates, the optimal decoder from $\Psi$ has essentially the same
performance as the optimal decoder from $\Phi^{*}$. Therefore, as
was mentioned in Subsection \ref{sub:System-Model-and}, using a single,
properly-optimized, threshold function for all the codebooks in the
ensemble may not incur a loss in exponents at all. Finally, we remark
that similar behavior was observed for a wide range of $T$. 


\appendices{\numberwithin{equation}{section}}

\section{\label{sec:Proofs}}
\begin{IEEEproof}[Proof of Proposition \ref{prop: simplified optimal decoder}]
The continuity of the error- and list size exponents in $T$, evident
from \eqref{eq: error exponent optimal} \eqref{eq:list exponent optimal},
implies that if $e^{nT}$ in \eqref{eq: optimal decoding regions Forney}
is replaced by any sequence $\tau_{n}\doteq e^{nT}$, then the same
exponents are achieved. Also, as the number of possible joint types
is polynomial in $n$ \cite[Chapter 2]{Csiszar2011}, we have 
\[
K_{m,n}\dfn\frac{\max_{Q}N_{m}(Q|\mathbf{y})\cdot e^{nf(Q)}}{\sum_{Q}N_{m}(Q|\mathbf{y})\cdot e^{nf(Q)}}\doteq1
\]
for all $1\leq m\leq M$. Thus, ${\cal R}_{m}^{*}$ and the decoding
sets 
\begin{align*}
\left\{ \mathbf{y}:P(\mathbf{y}|\mathbf{x}_{m})\geq e^{nT}\cdot K_{m,n}\cdot\sum_{m=1}^{M}P(\mathbf{y}|\mathbf{x}_{m})\right\}  & =\\
\left\{ \mathbf{y}:e^{nf(\hat{Q}_{\mathbf{x}_{m}\mathbf{y}})}\geq e^{nT}\cdot K_{m,n}\cdot\sum_{Q}N_{m}(Q|\mathbf{y})\cdot e^{nf(Q)}\right\}  & =\\
\left\{ \mathbf{y}:e^{nf(\hat{Q}_{\mathbf{x}_{m}\mathbf{y}})}\geq e^{nT}\cdot\max_{Q}N_{m}(Q|\mathbf{y})\cdot e^{nf(Q)}\right\}  & =\tilde{{\cal R}}_{m}
\end{align*}
have the same random coding exponents.
\end{IEEEproof}

\begin{IEEEproof}[Proof of Theorem \ref{thm:exp threshold exponents}]
We assume, without loss of generality, that $\mathbf{x}_{1}$ was
transmitted. For a random codebook, the error probability is 
\begin{align}
\overline{P_{e}({\cal C},\phi)} & =\sum_{\mathbf{x}_{1},\mathbf{y}}P(\mathbf{x}_{1},\mathbf{y})\cdot\P(\mbox{error}|\mathbf{x}_{1},\mathbf{y})\\
 & \trre[=,a]\sum_{\mathbf{x}_{1},\mathbf{y}}P(\mathbf{x}_{1},\mathbf{y})\cdot\P\left\{ f(\tilde{Q})<\max_{m>1}h(\hat{Q}_{\mathbf{X}_{m}\mathbf{y}})\right\} \\
 & =\sum_{\mathbf{x}_{1},\mathbf{y}}P(\mathbf{x}_{1},\mathbf{y})\P\bigcup_{m>1}\left\{ f(\tilde{Q})<h(\hat{Q}_{\mathbf{X}_{m}\mathbf{y}})\right\} \\
 & \trre[\doteq,b]\sum_{\mathbf{x}_{1},\mathbf{y}}P(\mathbf{x}_{1},\mathbf{y})\cdot\min\left\{ M\cdot\P\left\{ f(\tilde{Q})<\exp\left[n\cdot h(\hat{Q}_{\mathbf{X}_{2}\mathbf{y}})\right]\right\} ,1\right\} ,
\end{align}
where in $(a)$ we have introduced the notation $\tilde{Q}\dfn\hat{Q}_{\mathbf{x}_{1}\mathbf{y}}$,
and in $(b)$ we have used the exponential tightness of the union
bound (limited by unity) for a union of \emph{exponential} number
of events, assuming that they are pairwise independent \cite[Lemma A.2]{shulman2003communication}.
Now, from the method of types, 
\begin{align*}
\P\left\{ f(\tilde{Q})<\exp\left[n\cdot h(\hat{Q}_{\mathbf{X}_{2}\mathbf{y}})\right]\right\}  & =\sum_{Q:\: Q_{Y}=\tilde{Q}_{Y},h(Q)\geq f(\tilde{Q})}e^{-n\cdot I(Q)}\\
 & \doteq\exp\left[-\min_{Q:\: Q_{Y}=\tilde{Q}_{Y},h(Q)\geq f(\tilde{Q})}I(Q)\right]
\end{align*}
and then, it is easy to show that the resulting exponent is as in
\eqref{eq: error exponent exponential threshold}. 

Next, let us evaluate the random coding list size exponent 
\begin{align}
\overline{L({\cal C},\phi)} & =\sum_{\mathbf{x}_{1},\mathbf{y}}P(\mathbf{x}_{1},\mathbf{y})\cdot\E\left[L|\mathbf{X}_{1}=\mathbf{x}_{1},\mathbf{Y}=\mathbf{y}\right].\label{eq: list size averaging x1_y}
\end{align}
Given $\mathbf{x}_{1},\mathbf{y}$ (with $\hat{Q}_{\mathbf{x}_{1}\mathbf{y}}=\tilde{Q}$)
we have 
\begin{align*}
\E\left[L|\mathbf{X}_{1}=\mathbf{x}_{1},\mathbf{Y}=\mathbf{y}\right] & =\E\left[\sum_{m=2}^{M}\I\left\{ P(\mathbf{y}|\mathbf{X}_{m})\geq\max\left\{ \max_{l>1,l\neq m}\exp\left[n\cdot h(\hat{Q}_{\mathbf{X}_{l}\mathbf{y}})\right],\exp\left[n\cdot h(\hat{Q}_{\mathbf{x}_{1}\mathbf{y}})\right]\right\} \right\} \right]\\
 & =\sum_{m=2}^{M}\P\left\{ P(\mathbf{y}|\mathbf{X}_{m})\geq\max\left\{ \max_{l>1,l\neq m}\exp\left[n\cdot h(\hat{Q}_{\mathbf{X}_{l}\mathbf{y}})\right],\exp\left[n\cdot h(\hat{Q}_{\mathbf{x}_{1}\mathbf{y}})\right]\right\} \right\} \\
 & \doteq e^{nR}\cdot\P\left\{ e^{n\cdot f(\hat{Q}_{\mathbf{X}_{2}\mathbf{y}})}\geq\max\left\{ \max_{l>2}e^{n\cdot h(\hat{Q}_{\mathbf{X}_{l}\mathbf{y}})},e^{n\cdot h(\hat{Q}_{\mathbf{x}_{1}\mathbf{y}})}\right\} \right\} \\
 & \trre[=,a]e^{nR}\cdot\sum_{Q'}\P(\hat{Q}_{\mathbf{X}_{2}\mathbf{y}}=Q')\cdot\P\left\{ e^{n\cdot f(Q')}\geq\max\left\{ \max_{l>2}e^{n\cdot h(\hat{Q}_{\mathbf{X}_{l}\mathbf{y}})},e^{n\cdot h(\tilde{Q})}\right\} \right\} \\
 & =e^{nR}\cdot\sum_{Q':\: f(Q')\geq h(\tilde{Q})}\P(\hat{Q}_{\mathbf{X}_{2}\mathbf{y}}=Q')\cdot\P\left\{ e^{n\cdot f(Q')}\geq\max_{l>2}e^{n\cdot h(\hat{Q}_{\mathbf{X}_{l}\mathbf{y}})}\right\} \\
 & =e^{nR}\cdot\sum_{Q':\: f(Q')\geq h(\tilde{Q})}\P(\hat{Q}_{\mathbf{X}_{2}\mathbf{y}}=Q')\cdot\P\bigcap_{l>2}\left\{ e^{n\cdot h(\hat{Q}_{\mathbf{X}_{l}\mathbf{y}})}\leq e^{n\cdot f(Q')}\right\} 
\end{align*}
where $(a)$ is using the fact that the codewords are drawn independently
and with the same distribution, and the law of total probability.
Letting $\tilde{Q}_{l}$ be the joint type of $(\mathbf{X}_{l},\mathbf{y})$,
for any given $u\in\reals$ 
\begin{align*}
\P\bigcap_{l>2}\left\{ e^{n\cdot h(\tilde{Q}_{l})}\leq e^{nu}\right\}  & =\left[\P\left\{ e^{n\cdot h(\tilde{Q}_{l})}\leq e^{nu}\right\} \right]^{M-2}\\
 & =\left[1-\P\left\{ \tilde{Q}_{l}:\: h(\tilde{Q}_{l})>u\right\} \right]^{M-2}\\
 & \doteq\left[1-e^{-n\cdot\mathbf{U}(\tilde{Q}_{Y},u)}\right]^{e^{nR}}
\end{align*}
 where
\[
\mathbf{U}(\tilde{Q}_{Y},u)\dfn\min_{Q':\: Q'{}_{Y}=\tilde{Q}_{Y},h(Q')\geq u}I(Q').
\]
Then, 
\[
\P\bigcap_{l>2}\left\{ e^{n\cdot h(\tilde{Q}_{l})}\leq e^{nu}\right\} \doteq\begin{cases}
1, & R\leq\mathbf{U}(\tilde{Q}_{Y},u)\\
0, & R>\mathbf{U}(\tilde{Q}_{Y},u)
\end{cases}
\]
and so 
\begin{align*}
\E\left[L|\mathbf{X}_{1}=\mathbf{x}_{1},\mathbf{Y}=\mathbf{y}\right] & \doteq e^{nR}\cdot\sum_{Q':\: f(Q')\geq h(\tilde{Q})}\P(\hat{Q}_{\mathbf{X}_{2}\mathbf{y}}=Q')\cdot\I\left\{ \mathbf{U}(\tilde{Q}_{Y},f(Q'))\geq R\right\} \\
 & \doteq\exp\left[-n\cdot\left(\min_{Q\in{\cal D}(\tilde{Q})}I(Q)-R\right)\right]
\end{align*}
where 
\[
{\cal D}(\tilde{Q})\dfn\left\{ Q:Q_{Y}=\tilde{Q}_{Y},\mathbf{U}(\tilde{Q}_{Y},f(Q))\geq R,f(Q)\geq h(\tilde{Q})\right\} .
\]
To simplify the set ${\cal D}(\tilde{Q})$, notice that the condition
$\mathbf{U}(\tilde{Q}_{Y},f(Q))\geq R$ is actually
\[
\forall Q':Q'{}_{Y}=\tilde{Q}_{Y},h(Q')\geq f(Q)\:\Rightarrow\: I(Q')\geq R
\]
and equivalent to 

\[
\forall Q':Q'_{Y}=\tilde{Q}_{Y},I(Q')<R\:\Rightarrow\: h(Q')<f(Q).
\]
Thus, from continuity, the set ${\cal D}(\tilde{Q})$ can equivalently
be written as in \eqref{eq: cal=00007BD=00007D definition}, where
$\mathbf{V}(Q_{Y},R)$ is as defined in \eqref{eq: V_bold definition}.
After averaging w.r.t. $(\mathbf{x}_{1},\mathbf{y})$ as in \eqref{eq: list size averaging x1_y},
the list size exponent \eqref{eq: list exponent exponential threshold}
is obtained.
\end{IEEEproof}

\begin{IEEEproof}[Proof of Corollary \ref{cor:output exponents}]
We use the general expression for the exponents of a decoder $\phi_{h}\in\Psi$,
and obtain the exponents of $\phi_{g}\in\Lambda_{1}$ by setting $h(Q)=g(Q_{Y})$.
For the error exponent, we get 
\begin{align*}
E_{e}(R,g) & =\min_{\tilde{Q}}\min_{Q:\: Q_{Y}=\tilde{Q}_{Y},g(Q_{Y})\geq f(\tilde{Q})}\left\{ D(\tilde{Q}_{Y|X}||W|P_{X})+\left[I(Q)-R\right]_{+}\right\} \\
 & =\min_{Q}\min_{\tilde{Q}:\: Q_{Y}=\tilde{Q}_{Y},g(Q_{Y})\geq f(\tilde{Q})}\left\{ D(\tilde{Q}_{Y|X}||W|P_{X})+\left[I(Q)-R\right]_{+}\right\} \\
 & =\min_{Q}\min_{\tilde{Q}:\: Q_{Y}=\tilde{Q}_{Y},g(\tilde{Q}_{Y})\geq f(\tilde{Q})}\left\{ D(\tilde{Q}_{Y|X}||W|P_{X})+\left[I(Q)-R\right]_{+}\right\} \\
 & =\min_{Q'_{Y}}\min_{Q:\: Q_{Y}=Q'}\min_{\tilde{Q}:\:\tilde{Q}_{Y}=Q'_{Y},g(\tilde{Q}_{Y})\geq f(\tilde{Q})}\left\{ D(\tilde{Q}_{Y|X}||W|P_{X})+\left[I(Q)-R\right]_{+}\right\} \\
 & =\min_{Q'_{Y}}\min_{\tilde{Q}:\:\tilde{Q}_{Y}=Q'_{Y},g(\tilde{Q}_{Y})\geq f(\tilde{Q})}\left\{ D(\tilde{Q}_{Y|X}||W|P_{X})+\min_{Q:\: Q_{Y}=Q'_{Y}}\left[I(Q)-R\right]_{+}\right\} \\
 & =\min_{Q'_{Y}}\min_{\tilde{Q}:\:\tilde{Q}_{Y}=Q'_{Y},g(\tilde{Q}_{Y})\geq f(\tilde{Q})}D(\tilde{Q}_{Y|X}||W|P_{X})
\end{align*}
which is \eqref{eq: error exponent output}. For the list size exponent,
we first obtain 
\[
{\cal D}(\tilde{Q})=\left\{ Q:Q_{Y}=\tilde{Q}_{Y},f(Q)>\max\left\{ \mathbf{V}(Q_{Y},R),g(Q_{Y})\right\} \right\} ,
\]
where
\[
\mathbf{V}(Q_{Y},R)=\max_{Q':\: Q'{}_{Y}=Q_{Y},I(Q')\leq R,}g(Q'{}_{Y})=g(Q'_{Y})
\]
and the last equality is due to the feasibility of the maximization,
when setting $Q'=P_{X}\times Q_{Y}$. So, 
\[
{\cal D}(\tilde{Q})=\left\{ Q:Q_{Y}=\tilde{Q}_{Y},f(Q)>g(\tilde{Q}_{Y})\right\} 
\]
and \eqref{eq: list exponent exponential threshold} implies \eqref{eq: list exponent output}. 
\end{IEEEproof}

\begin{IEEEproof}[Proof of Corollary \ref{cor:ML exponents}]
Follows directly by substituting $h(Q)=T+f(Q)$ in \eqref{eq: error exponent exponential threshold}
and \eqref{eq: list exponent exponential threshold}.
\end{IEEEproof}

\begin{IEEEproof}[Proof of Theorem \ref{thm: optimal threshold output}]
From \eqref{eq: error exponent output}, if $g(Q_{Y})$ is such that
$E_{e}(R,g)>\EE$ then equivalently, the following condition holds:
\[
D(Q_{Y|X}||W|P_{X})\leq\EE\Rightarrow g(Q_{Y})\leq f(Q).
\]
Clearly, under this requirement on $E_{e}(R,g)$, the threshold $g(Q_{Y})$
should be chosen as large as possible in order to maximize the list
size exponent. Thus, the optimal (maximal) threshold function is given
by \eqref{eq: g_star definition}. The resulting list size exponent
is immediate from eq. \eqref{eq: list exponent output}.
\end{IEEEproof}

\begin{IEEEproof}[Proof of Lemma \ref{lem: Optimal output threshold properties}]
The first two properties are straightforward to prove. For convexity
in $Q_{Y}$, first, since $f(Q)$ is linear in $Q$, the minimizer
$Q^{*}$ in $g^{*}(Q_{Y},\EE)$ always achieves the divergence constraint
with an equality. Second, let $Q_{Y}^{0}$ and $Q_{Y}^{1}$ be two
$Y$-marginals, and consider 
\[
Q_{Y}^{\alpha}=(1-\alpha)\cdot Q_{Y}^{0}+\alpha\cdot Q_{Y}^{1}.
\]
Also, let $Q_{0}^{*}$ and $Q_{1}^{*}$ be the corresponding minimizers
in $g^{*}(Q_{Y}^{0},\EE)$ and $g^{*}(Q_{Y}^{1},\EE)$, respectively.
Now, since for any $\alpha\in(0,1)$ the $Y$-marginal of 
\[
Q_{\alpha}\dfn(1-\alpha)\cdot Q_{0}^{*}+\alpha\cdot Q_{1}^{*}
\]
is exactly $Q_{Y}^{\alpha}$, and because the divergence is a convex
function then 
\[
D(Q_{\alpha,Y|X}||W|P_{X})\leq\EE,
\]
we obtain
\begin{align*}
g^{*}(Q_{Y}^{\alpha},\EE) & \leq f(Q_{\alpha})\\
 & \trre[=,a](1-\alpha)\cdot f(Q_{0}^{*})+\alpha\cdot f(Q_{1}^{*}).\\
 & =(1-\alpha)\cdot g^{*}(Q_{Y}^{0},\EE)+\alpha\cdot g^{*}(Q_{Y}^{1},\EE).
\end{align*}
where $(a)$ is due to linearity of $f(Q)$. This proves convexity.
\end{IEEEproof}

\begin{IEEEproof}[Proof of Theorem \ref{thm: optimal threshold unified}]
 Assume that $E_{e}(R,h)\geq\EE$ for a decoder $\phi_{h}\in\Psi$.
Under this requirement on $E_{e}(R,h)$ the threshold $h(Q)$ should
be chosen as large as possible in order to maximize the list size
exponent. Suppose we are given a joint type $Q$, and notice that
the requirement $E_{e}(R,h)>\EE$ is equivalent to 
\[
\forall\tilde{Q}:Q_{Y}=\tilde{Q}_{Y},D(\tilde{Q}_{Y|X}||W|P_{X})+[I(Q)-R]_{+}\leq\EE\:\Rightarrow\: h^{*}(Q,R,\EE)<f(\tilde{Q})
\]
or

\[
h^{*}(Q,R,\EE)\leq\min_{\tilde{Q}:\:\tilde{Q}_{Y}=Q_{Y},D(\tilde{Q}_{Y|X}||W|P_{X})+[I(Q)-R]_{+}\leq\EE}f(\tilde{Q})
\]
we obtain \eqref{eq: optimal threhsold exponential}. 

For the optimal threshold function $h^{*}(Q,R,\EE)$, the resulting
error exponent is $\EE$ by assumption. Let us find now the achieved
$E_{l}^{*}(\Psi,R,\EE)$ given by
\begin{align}
E_{l}^{*}(\Psi,R,\EE) & =\min_{\tilde{Q}}\min_{Q\in\hat{{\cal D}}(\tilde{Q},R,\EE)}\left\{ D(\tilde{Q}_{Y|X}||W|P_{X})+I(Q)-R\right\} \label{eq: exponent list optimal exponential threshold optimization}
\end{align}
where
\[
\hat{{\cal D}}(\tilde{Q},R,\EE)\dfn\left\{ Q:Q_{Y}=\tilde{Q}_{Y},f(Q)\geq\max\left\{ \mathbf{V}^{*}(Q_{Y},R,\EE),h^{*}(\tilde{Q},R,\EE)\right\} \right\} ,
\]
and in this case 
\begin{align*}
\mathbf{V}^{*}(Q_{Y},R,\EE) & \dfn\max_{Q':\: Q'{}_{Y}=Q_{Y},I(Q')\leq R,}h^{*}(Q',R,\EE)\\
 & =g^{*}(Q_{Y},\EE)
\end{align*}
The set $\hat{{\cal D}}(\tilde{Q},R,\EE)$ can be simplified to the
set ${\cal D}^{*}(\tilde{Q},R,\EE)$ in \eqref{eq: D* for optimal exponential threshold},
by showing that $\mathbf{V}^{*}(Q_{Y},R,\EE)$ is never strictly larger
than $h^{*}(\tilde{Q},R,\EE)$. This can be verified by separating
the outer minimization over $\tilde{Q}$ in \eqref{eq: exponent list optimal exponential threshold optimization},
into two cases, which both satisfy $\mathbf{V}^{*}(Q_{Y},R,\EE)\leq h^{*}(\tilde{Q},R,\EE)$.
For $I(\tilde{Q})\leq R$ 
\begin{align*}
\mathbf{V}^{*}(\tilde{Q}_{Y},R,\EE) & =g^{*}(\tilde{Q}_{Y},\EE)\\
 & =h^{*}(\tilde{Q},R,\EE),
\end{align*}
and for $I(\tilde{Q})>R$, 
\begin{align*}
\mathbf{V}^{*}(\tilde{Q}_{Y},R,\EE) & =g^{*}(\tilde{Q}_{Y},\EE)\\
 & \leq g^{*}(\tilde{Q}_{Y},\EE-I(\tilde{Q})+R)
\end{align*}
using Lemma \ref{lem: Optimal output threshold properties} (property
\ref{enu: output optimal non increasing}). Thus, we obtain \eqref{eq: list exponent optimal exponential threshold proof}.
\end{IEEEproof}

\begin{IEEEproof}[Proof of Theorem \ref{thm: optimal threshold ML}]
Here, the requirement $E_{e}(R,T)\geq\EE$ is equivalent to
\[
D(\tilde{Q}_{Y|X}||W|P_{X})+[I(Q)-R)]_{+}\leq\EE\Rightarrow T<f(\tilde{Q})-f(Q)
\]
which leads to \eqref{eq: optimal threhsold ML} using
\begin{align*}
T^{*}(R,\EE) & =\min_{\tilde{Q}}\min_{Q:\:\tilde{Q}_{Y}=Q_{Y},D(\tilde{Q}_{Y|X}||W|P_{X})+[I(Q)-R]_{+}\leq\EE}\left\{ f(\tilde{Q})-f(Q)\right\} \\
 & =\min_{Q}\min_{\tilde{Q}:\:\tilde{Q}_{Y}=Q_{Y},D(\tilde{Q}_{Y|X}||W|P_{X})\leq\EE-[I(Q)-R]_{+}}\left\{ f(\tilde{Q})-f(Q)\right\} \\
 & =\min_{Q}\left\{ g^{*}(Q_{Y},\EE-[I(Q)-R]_{+})-f(Q)\right\} \\
 & =\min_{Q}\left\{ h^{*}(Q,R,\EE)-f(Q)\right\} .
\end{align*}
The list size exponent \eqref{eq: list size optimal exponent ML threshold}
is immediate.
\end{IEEEproof}

\begin{IEEEproof}[Proof of Proposition \ref{prop: optimality of output threshold}]
We have
\begin{align*}
E_{l}^{*}(\Psi,R,\EE) & =\min_{\tilde{Q}}\min_{Q\in{\cal D}^{*}(\tilde{Q},R,\EE)}\left\{ D(\tilde{Q}_{Y|X}||W|P_{X})+I(Q)-R\right\} \\
 & \leq\min_{\tilde{Q}:\: I(\tilde{Q})\leq R}\min_{Q:\: Q_{Y}=\tilde{Q}_{Y},f(Q)>g^{*}(Q_{Y},\EE)}\left\{ D(\tilde{Q}_{Y|X}||W|P_{X})+I(Q)-R\right\} \\
 & \trre[=,a]\min_{\tilde{Q}}\min_{Q:\: Q_{Y}=\tilde{Q}_{Y},f(Q)>g^{*}(Q_{Y},\EE)}\left\{ D(\tilde{Q}_{Y|X}||W|P_{X})+I(Q)-R\right\} \\
 & =E_{l}^{*}(\Lambda_{1},R,\EE).
\end{align*}
where $(a)$ is for $R>\overline{R}_{\s[cr]}(\EE)$. Also, since $\Lambda_{1}\subset\Psi$,
we have $E_{l}^{*}(\Psi,R,\EE)\geq E_{l}^{*}(\Lambda_{1},R,\EE)$
and so equality is obtained.
\end{IEEEproof}

\begin{IEEEproof}[Proof of Proposition \ref{prop: optimality of ML threshold}]
For $\phi_{T}^{*}$ and $R\leq\underline{R}_{\s[cr]}(T)$ 
\begin{align*}
E_{e}^{*}(R,T) & \leq E_{a}(R,T)\\
 & =\min_{\tilde{Q}}\min_{Q:\: Q_{Y}=\tilde{Q}_{Y},f(Q)+T\geq f(\tilde{Q}),I(Q)\geq R}D(\tilde{Q}_{Y|X}||W|P_{X})+I(Q)-R\\
 & =D(\tilde{Q}_{Y|X}^{*}||W|P_{X})+I(Q^{*})-R,
\end{align*}
and also $E_{l}^{*}(R,T)=E_{e}^{*}(R,T)+T$. 

Now, we consider the optimization problem
\begin{equation}
\min_{\tilde{Q}}\min_{Q:\: Q_{Y}=\tilde{Q}_{Y}}\left\{ D(\tilde{Q}_{Y|X}||W|P_{X})+I(Q)\right\} .\label{eq: error exponent ML R=00003D0, T=00003D0 no likelihood constraint}
\end{equation}
and show that its solution, which we denote by $(\tilde{Q}_{0},Q_{0})$,
satisfies $\tilde{Q}_{0}=Q_{0}$. To see this, we utilize the identity
\begin{equation}
D(\tilde{Q}_{Y|X}||W|P_{X})+I(Q)=D(Q_{Y|X}||W|P_{X})+I(\tilde{Q})+f(Q)-f(\tilde{Q})\label{eq: identity}
\end{equation}
which holds under the assumption $Q_{Y}=\tilde{Q}_{Y}$, and can be
proved using simple algebraic manipulations. Now, for any given $(Q,\tilde{Q})$
such that $Q_{Y}=\tilde{Q}_{Y}$
\begin{alignat}{1}
D(\tilde{Q}_{Y|X}||W|P_{X})+I(Q) & \trre[=,a]\frac{1}{2}\left(D(\tilde{Q}_{Y|X}||W|P_{X})+D(Q_{Y|X}||W|P_{X})+I(Q)+I(\tilde{Q})\right)\nonumber \\
 & +\frac{1}{2}\left(f(Q)-f(\tilde{Q})\right)\\
 & \trre[\geq,b]D\left(\frac{1}{2}(\tilde{Q}_{Y|X}+Q_{Y|X})||W|P_{X}\right)+I\left(\frac{1}{2}(\tilde{Q}+Q)\right)+\frac{1}{2}\left(f(Q)-f(\tilde{Q})\right)\\
 & \trre[=,c]D\left(\frac{1}{2}(\tilde{Q}_{Y|X}+Q_{Y|X})||W|P_{X}\right)+I\left(\frac{1}{2}(\tilde{Q}+Q)\right)+\frac{1}{2}\left(f(Q)+f(\tilde{Q})\right)\nonumber \\
 & -f(\tilde{Q})\\
 & \trre[\geq,d]D\left(\frac{1}{2}(\tilde{Q}_{Y|X}+Q_{Y|X})||W|P_{X}\right)+I\left(\frac{1}{2}(\tilde{Q}+Q)\right)+\frac{1}{2}\left(f(Q)+f(\tilde{Q})\right)
\end{alignat}
where $(a)$ is because the right hand side of \eqref{eq: identity}
equals the average of both sides of \eqref{eq: identity}, $(b)$
is due to convexity of both the divergence and the mutual information%
\footnote{Indeed, when both marginals of $Q$ are constrained, $I(Q)=D(Q||P_{X}\times Q_{Y})$
and so convexity of the mutual information is implied by the convexity
of the divergence.%
}, $(c)$ is due to the linearity of $f(Q)$, and $(d)$ is due to
the negativity of $f(Q)$. Equalities are obtained in $(b)$ and $(d)$
by choosing $Q=\tilde{Q}$. Thus, 
\[
\min_{\tilde{Q}}\min_{Q:\: Q_{Y}=\tilde{Q}_{Y}}\left\{ D(\tilde{Q}_{Y|X}||W|P_{X})+I(Q)\right\} =\min_{Q}\left\{ D(Q_{Y|X}||W|P_{X})+I(Q)\right\} ,
\]
and so $\tilde{Q}_{0}=Q_{0}$. This implies that for $T\geq0$, $\tilde{Q}^{*}=Q^{*}=\tilde{Q}_{0}=Q_{0}$.
Thus, for $R\leq\underline{R}_{\s[cr]}(T)$ 
\begin{align*}
E_{e}(R,T) & =\min_{\tilde{Q}}\min_{Q:\: Q_{Y}=\tilde{Q}_{Y},f(Q)+T\geq f(\tilde{Q})}D(\tilde{Q}_{Y|X}||W|P_{X})+\left[I(Q)-R\right]_{+}\\
 & =D(\tilde{Q}_{Y|X}^{*}||W|P_{X})+I(Q^{*})-R\\
 & =E_{a}(R,T)\\
 & \geq E_{e}^{*}(R,T)
\end{align*}
On the other hand, for the list size exponent, 
\begin{align*}
E_{l}(R,T) & =\min_{\tilde{Q}}\min_{Q\in\underline{{\cal D}}(\tilde{Q},R,T)}\left\{ D(\tilde{Q}_{Y|X}||W|P_{X})+I(Q)-R\right\} \\
 & \trre[\geq,a]\min_{\tilde{Q}}\min_{Q:\: Q_{Y}=\tilde{Q}_{Y},f(Q)\geq f(\tilde{Q})+T}\left\{ D(\tilde{Q}_{Y|X}||W|P_{X})+I(Q)-R\right\} \\
 & \trre[\geq,b]\min_{\tilde{Q}}\min_{Q:\: Q_{Y}=\tilde{Q}_{Y},f(Q)\geq f(\tilde{Q})+T}\left\{ D(\tilde{Q}_{Y|X}||W|P_{X})+I(Q)-R+f(\tilde{Q})+T-f(Q)\right\} \\
 & \trre[=,c]\min_{\tilde{Q}}\min_{Q:\: Q_{Y}=\tilde{Q}_{Y},f(Q)\geq f(\tilde{Q})+T}\left\{ D(Q_{Y|X}||W|P_{X})+I(\tilde{Q})-R+T\right\} \\
 & =\min_{\tilde{Q}}\min_{Q:\: Q_{Y}=\tilde{Q}_{Y},f(\tilde{Q})\geq f(Q)+T}\left\{ D(\tilde{Q}_{Y|X}||W|P_{X})+I(Q)-R+T\right\} \\
 & \geq\min_{\tilde{Q}}\min_{Q:\: Q_{Y}=\tilde{Q}_{Y}}\left\{ D(\tilde{Q}_{Y|X}||W|P_{X})+I(Q)-R+T\right\} \\
 & =D(\tilde{Q}_{Y|X}^{*}||W|P_{X})+I(Q^{*})-R+T\\
 & =E_{e}(R,T)+T
\end{align*}
where inequality $(a)$ is obtained by removing the constraint $f(Q)\geq\underline{\mathbf{V}}(Q_{Y},R)$
from the set $\underline{{\cal D}}(\tilde{Q},R,T)$, inequality $(b)$
is using the constraint $f(Q)\geq f(\tilde{Q})+T$. Equality $(c)$
is using the identity \eqref{eq: identity}, and $(d)$ is simply
a substitution of notation $Q\leftrightarrow\tilde{Q}$. To conclude,
we have obtained $E_{e}(R,T)\geq E_{e}^{*}(R,T)$ and $E_{l}(R,T)=E_{e}(R,T)+T\geq E_{e}^{*}(R,T)+T=E_{l}^{*}(R,T)$.
Since $\phi_{T}^{*}$ provides the optimal trade-off between the error
exponent and list size exponent, we obtain the desired result. 
\end{IEEEproof}

\section{\label{sec:Justification-of-the}}

In \cite{exact_erasure}, random coding exponents were derived for
an optimal decoder in the erasure mode, i.e. with $T\geq0$, and expressions
for both $E_{e}^{*}(R,T)$ and $E_{l}^{*}(R,T)$ were derived independently
(in the erasure case, $E_{e}(R,\phi)$ and $E_{l}(R,\phi)$ represent
the probability of erasure, and the probability of undetected error,
respectively). The reason for restricting the analysis to the erasure
mode is that the fact that the decoding regions overlap in the list
mode complicates analysis. Nonetheless, it can be easily verified
that this restriction is only needed for the analysis of $E_{l}^{*}(R,T)$,
and so that analysis of $E_{e}^{*}(R,T)$ is valid for this list mode
$T<0$. Moreover, it can be shown that $E_{l}^{*}(R,T)=E_{e}^{*}(R,T)+T$
must be satisfied in general (which was shown in \cite{exact_erasure}
only for the BSC), see \cite[Lemma 1]{universal_erasure_exact}. 

In addition, the random ensemble considered in \cite{exact_erasure}
is the i.i.d. ensemble over input distribution $P_{X}$, but the analysis
may be easily adapted to fixed composition codes with the same input
distribution, which may only lead to larger exponents. Basically,
divergence terms $D(Q_{X}||P_{X})$, where $Q_{X}$ is a generic distribution,
are omitted from exponential assessment of probabilities, and thus
from final expressions.

Therefore, from the above reasoning, the results of \cite{exact_erasure}
are also applicable here, with proper adjustments.

It should also be noted that there is an error at the end of the proof
of \cite[Theorem 1]{exact_erasure}, where it was claimed that $\min\{E_{a}(R,T),E_{b}(R,T)\}$,
which may not be true in general. The correct expression is as in
\eqref{eq: error exponent optimal}.

\bibliographystyle{plain}
\bibliography{Threshold_Decoding}

\end{document}